\documentclass{ws-procs9x6}
\usepackage{amssymb,amsfonts,float}
\usepackage{amsmath}

\newcommand{\beq}{\begin{eqnarray}}
\newcommand{\eeq}{\end{eqnarray}}

\begin{document}

\title{Features of $SU(N)$ Gauge Theories\footnote{\uppercase{S}ummary of the
talk presented by \uppercase{B}. \uppercase{L}ucini and of the poster
presented by \uppercase{U}. \uppercase{W}enger.} \footnote{\uppercase{W}ork
partially supported by the \uppercase{R}oyal \uppercase{S}ociety and
the \uppercase{L}ockey \uppercase{C}ommittee.}}

\author{B. Lucini, M. Teper and U. Wenger}

\address{Theoretical Physics, 
University of Oxford,\\
1 Keble Road,
Oxford OX1 3NP, United Kingdom\\
E-mail: lucini,teper,wenger@thphys.ox.ac.uk}  

\maketitle

\abstracts
{We review recent lattice results for the large $N$
limit of $SU(N)$ gauge theories. In particular, we focus on
glueball masses, topology and its relation to chiral symmetry
breaking (relevant for
phenomenology), on the tension of strings connecting sources in higher
representations of the gauge group (relevant for models of confinement
and as a comparative ground for theories beyond the Standard Model) and on 
the finite temperature deconfinement phase transition (relevant for RHIC-like
experiments). In the final part we present open challenges for the future.}

\section{Introduction}
The possibility that observables in $SU(N)$ gauge theories
are (at least for large enough $N$) smooth functions of $N$ has been
advocated a long time ago\cite{largeN}. In particular, it has been proven
diagrammatically
that in the limit $N \to \infty$ the theory is simpler (only planar
diagrams survive) and that in a
neighbourhood of that limit the leading corrections due to a finite $N$
go as $1/N^2$. This can have practical implications for our understanding
of QCD if the gauge group of this theory, $SU(3)$, shares the bulk of
the physics with $SU(\infty)$. To verify if this is the case, an investigation
from first principles is mandatory.
Another motivation for studying the physics of $SU(N)$ at large $N$
within the conventional gauge theory
comes from calculations performed in ``beyond the Standard Model'' frameworks:
the bridge between the two approaches is often $SU(\infty)$
Yang-Mills\cite{reviewstring}.\\
Lattice calculations are the most reliable tool for
investigations of gauge theories from first principles.
In the following we give a quick overview of recent lattice results obtained
by our group for $SU(\infty)$. For more details
about the calculations we refer to the quoted literature.

\section{The lowest-lying masses in the spectrum}
\begin{figure}[ht]
\begin{center}
\includegraphics[scale=0.3,angle=270]{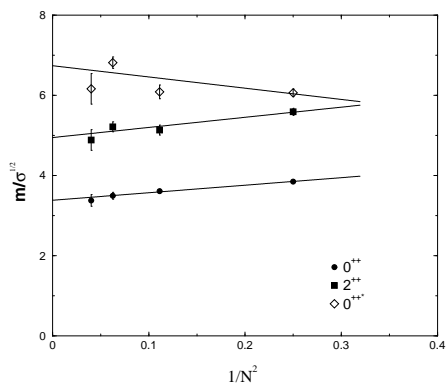} 
\caption{The mass of the lightest glueballs as a function of $1/N^2$. Solid
lines are our best fits to the data. \label{fig1}}
\end{center}
\end{figure}

We have studied $SU(N)$ gauge theories for $2 \le N \le 8$. At zero
temperature, the theory is confined at all these $N$, which suggests
that the large $N$ limit is confining. We use the value of the
string tension ${\sigma}$ to set the scale for the studied observables.\\
A large $N$ understanding of the mass spectrum can help in identifying
glueball states in experiments. We have looked at
the masses of the $0^{++}$, $2^{++}$ and $0^{++*}$ glueballs as a function of
$N$. We have measured those quantities for $2 \le N \le 5$.
Our results for $m/\sqrt{\sigma}$ are plotted in fig.~\ref{fig1} as a
function of $1/N^2$. The choice of the independent variable is dictated by the 
fact that the first expected correction to the large $N$ limit value
of an observable has this
functional form. A fit to the data using only this correction generally works
pretty well all the way down to $SU(2)$ ({\it i.e.} there is a somehow
unexpected precocious onset of the large $N$ behaviour) and allows us to
extract the value of the spectrum at $N=\infty$. We find\cite{blmt1} 
\beq
\begin{array}{l}
{m}/{\sqrt{\sigma}} =  3.341(76) + {1.75}/{N^2} \\
{m}/{\sqrt{\sigma}} =  4.93(13) + {2.58}/{N^2} \\
{m}/{\sqrt{\sigma}} =  6.48(35) - {1.7}/{N^2}
\end{array}
\eeq
respectively for the $0^{++}$, the $2^{++}$ and the $0^{++*}$
glueballs.

\section{The topological susceptibility}
Another interesting quantity in the large $N$ limit is the topological
susceptibility $\chi_t$,
which is related in that limit to the $\eta^{\prime}$ mass
by the Witten-Veneziano formula\cite{wvformula}. This formula is a good
approximation even when the $SU(3)$ value of the topological susceptibility
is plugged in. An explanation of this fact can be found by investigating how
$\chi_t$ varies with $N$. We find\cite{blmt1}
\beq 
\label{chi}
{\chi}_{t}^{1/4}/\sqrt{\sigma} = 0.3739(59) + {0.439}/{N^2} \ .
\eeq
The determination of $\chi_t$ at $N=\infty$ given in ref.~\cite{pisa2}
agrees with our result.

\section{Topology and chiral symmetry breaking$^\text{\normalfont a}$}
\footnotetext{$^\text{a}$We thank N.~Cundy who contributed to the
results in this section.}  From the Banks-Casher relation we know that
the chiral condensate is proportional to the density of small
eigenmodes $\lambda$ of the Dirac operator, $\langle \bar \psi
\psi\rangle \sim \lim_{\lambda \rightarrow 0} \rho(\lambda)$.  Hence a
possible scenario for spontaneous chiral symmetry breaking is to
assume that the non-vanishing density is due to exact zero modes of
the Dirac operator which, through their interaction,
are lifted away from zero yielding $\lim_{\lambda \rightarrow 0}
\rho(\lambda) \neq 0$. As a consequence the near-zero modes would
have a topological origin since they emerged from the
topological zero modes. A comparison of the topological 
content of near-zero modes and zero modes of the $SU(N)$ Dirac
operator can tell us qualitatively whether 
this scenario holds 
in the large $N$ limit\cite{wenger}.
\begin{figure}[ht]
\begin{center}
\includegraphics[scale=0.3,angle=270]{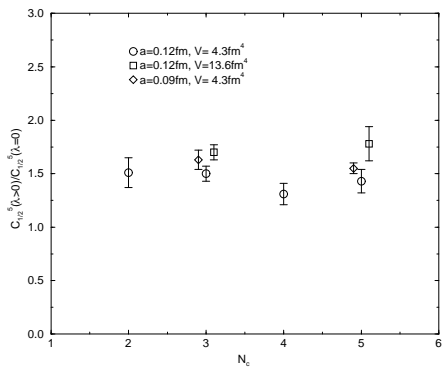} 
\caption{The normalised topological content of the near-zero modes
responsible for chiral symmetry breaking for different volumes $V$,
lattice spacings $a$ and $N$.\label{fig3}}
\end{center}
\end{figure}
We find that the topological and chiral contents of both the zero and
near-zero modes become smaller as $N$ increases, but at roughly the
same rate.  As a consequence the topological content of the near-zero
modes normalised by the one of the zero modes (fig.~\ref{fig3}) is
constant for all $N$. We find that this remains qualitatively true as
we vary the volume $V$ and the lattice spacing $a$ and we are
therefore able to conclude that topology indeed drives chiral symmetry
breaking for all $N$.

\section{$k$-strings}
For $N \ge 4$ strings connecting sources in higher representations of
the gauge group can be stable. This is a consequence of the
$\mathbb{Z}(N)$ symmetry of the confined phase.  Taking into account
also charge conjugation, it is easy to see that the number of stable
stringy states for $SU(N)$ is the integer part of $N/2$.  These states
are referred to in the literature as $k$-strings and the rank of the
representation of the sources is called N-ality. Within a class of
strings with the same N-ality the string tension is unique. We
indicate this tension by $\sigma_k$.\\ The {\em a priori} unknown
value of the ratio $\sigma _k/ \sigma$ poses constraints to effective
models of confinement. Moreover, it can shed light on the connection
between QCD and some ``beyond the Standard Model''
theories\cite{mqcd}. Usually in these frameworks one obtains the
so-called sine formula: \beq
\label{sine}
{\sigma_k}/{\sigma} = {\sin(k\pi/N)}/{\sin(\pi/N)} \ ; \eeq however,
this is not a universal feature\cite{konishi}.  The first lattice
calculation of this ratio was performed in \cite{wingate} for
$SU(4)$. Although a useful continuum extrapolation could not be
obtained, it was found that the $k=2$ string is a genuine bound
state. Recent calculations\cite{usstring,usstring2} have obtained a
precise continuum determination of ${\sigma_k}/{\sigma}$. This is
compatible with both the sine formula and the Casimir scaling ansatz,
which predicts that ${\sigma_k}/{\sigma}$ is equal to the ratio of the
lowest quadratic Casimir operators in the class of the representations
with the N-ality of the sources: \beq
\label{casimir}
\sigma_k/\sigma = k(N-k)/(N-1) \ .
\eeq
While the authors of \cite{pisa1} claim to be able to exclude Casimir scaling,
in our opinion the question is far from being settled: the numerical values
of~(\ref{sine})~and~(\ref{casimir}) are close enough for systematic effects
to become relevant. We are currently trying to deal with those issues.\\
Indirect information on $\sigma_k/\sigma$ can be extracted from the size
of the corresponding strings\cite{usstring2,shifman2} or from the behaviour
of the leading correction to its asymptotic value as $N$
increases\cite{shifman2,mikereview}.

\section{The deconfinement phase transition}
The quark-gluon plasma phase of QCD is currently being investigated
experimentally. Some of the related theoretical questions like the late
onset of the Stephan-Boltzman law can be answered in the context of large
$N$. As a first step in that direction, we have investigated the physics
of the deconfinement phase transition as $N$
varies and then we have extrapolated our results to the large $N$ limit.
For the deconfinement temperature $T_c$ we find\cite{blmtuw1,blmtuw2}
\beq
\label{extrnew}
T_c/\sqrt{\sigma} = 0.596(4) + 0.453(30)/N^2 \ .
\eeq
For $N \ge 3$ the transition is first order, with a monotonically increasing
latent heat as a function of $N$, which suggests that also for $SU(\infty)$
the transition is first order. In fact, an extrapolation of the latent heat
to $N = \infty$ using the leading expected $O(1/N^2)$ correction predicts a
finite value for this quantity in the limiting case.\\
\begin{figure}[ht]
\begin{center}
\includegraphics[scale=0.3,angle=270]{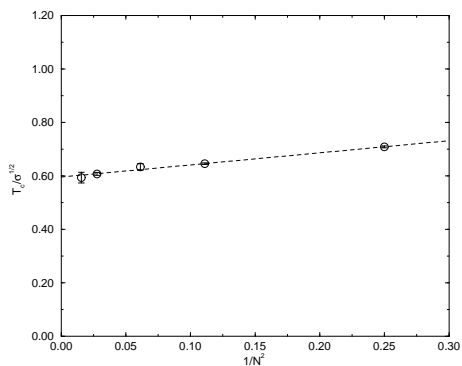} 
\caption{$T_c/\sqrt{\sigma}$ vs $1/N^2$. The solid
line is our best fit to the data.\label{fig2}}
\end{center}
\end{figure}
The interface tension between the confined and
the deconfined phase grows as $N$ is increased and seems to diverge in the
critical region at infinite $N$. Interpreted in terms of the Master
Field\cite{witten}, this gives rise to the speculation
that in the large $N$ limit there are several Master Fields separated by
infinite energy barriers\cite{blmtuw2}.
The rich physics of the limiting case is given by the interplay between those
vacua at large but finite $N$.

\section{Conclusions}
$SU(N)$ gauge theories have a sensible large $N$ limit that can be
studied by lattice techniques. The results can have relevant implications
for our understanding of QCD and of the physics beyond the Standard Model.\\
Possible future directions of our investigations include the physics of
confinement, the full mass spectrum of glueballs and the equation of
state at finite temperature.

\section*{Acknowledgements}
B.L. would like to thank P. Faccioli, J. Greensite, K. Konishi and M.
Shifman for interesting discussions. B.L. is supported by a EU Marie Curie
Fellowship and U.W. by a PPARC SPG Fellowship.

\end{document}